# A Way Memoization Technique for Reducing Power Consumption of Caches in Application Specific Integrated Processors

Tohru Ishihara     Farzan Fallah

Fujitsu Laboratories of America, Inc.
1240 East Arques Avenue, M/S 345, Sunnyvale, CA 94085 USA
{Toru.Ishihara, Farzan.Fallah}@us.fujitsu.com

**Abstract**

*This paper presents a technique for eliminating redundant cache-tag and cache-way accesses to reduce power consumption. The basic idea is to keep a small number of Most Recently Used (MRU) addresses in a Memory Address Buffer (MAB) and to omit redundant tag and way accesses when there is a MAB-hit. Since the approach keeps only tag and set-index values in the MAB, the energy and area overheads are relatively small even for a MAB with a large number of entries. Furthermore, the approach does not sacrifice the performance. In other words, neither the cycle time nor the number of executed cycles increases. The proposed technique has been applied to Fujitsu VLIW processor (FR-V) and its power saving has been estimated using NanoSim. Experiments for 32kB 2-way set associative caches show the power consumption of I-cache and D-cache can be reduced by 40% and 50%, respectively.*

## 1   Introduction

On-chip cache memories are one of the most power hungry components of microprocessors. For example, the on-chip caches of DEC Alpha 21164 dissipate 25% of the total power of the processor [1]. The StrongARM SA-110 processor, which specifically targets low power applications, dissipates about 43% of the power in its on-chip caches [2]. Thus, reducing the power consumption of a cache memory can greatly reduce the overall power consumption of a processor.

In [3-17], authors have proposed techniques which reduce the power consumption of on-chip cache memories. One simple approach is to employ a small L0-cache between a CPU core and its L1 cache, e.g., S-cache [4], block-Buffer [5], filter-cache [6], and loop-cache [7]. Since an L0-cache is small, it consumes less power per access. Therefore, if there is a hit in the L0-cache, the power consumption will be reduced. On the other hand, if there is a miss, one extra cycle is required to access the L1 cache. Another simple approach proposed is using a two-phase cache [8]. In the first phase, tags of all cache-ways are accessed to find the cache-way having the data. If there is a hit, in the second phase, only one of the cache-ways is activated. Although this approach can eliminate unnecessary way accesses, it results in a performance loss. The method proposed in [9] can also reduce the number of tag accesses by using a way-prediction table and accessing the tag and data of the predicted way only. This approach also involves a performance loss because in case of a misprediction, one extra cycle is required to perform tag comparison for all ways.

In this paper, we propose a new way memoization technique which eliminates redundant tag and way accesses to reduce the power consumption. The basic idea is to keep a small number of Most Recently Used (MRU) addresses in a *Memory Address Buffer* (MAB) and to omit redundant tag and way accesses when there is a MAB-hit. The MAB is accessed in parallel with the adder used for address generation (see Figure 1 and 2). The technique does not increase the delay of the circuit. Furthermore, this approach does not require modifying the cache architecture. This is considered an important advantage in industry because it makes it possible to use the processor core with previously designed caches or IPs provided by other vendors.

The rest of the paper is organized as follows. Section 2 describes related work on cache power reduction. Our approach which reduces the power consumption without any performance penalty, is presented in Section 3. Section 4 presents experimental results and discussion on the effectiveness of the approach, while Section 5 concludes the paper.

## 2   Related work

Panwar et al. have shown that cache-tag access and tag comparison do not need to be performed for all instruction fetches [4]. Consider an instruction $j$ which is executed immediately after an instruction $i$. There are four cases,

1. **Intra-cache-line sequential flow**
   This occurs when both $i$ and $j$ instructions reside on the same cache-line and $i$ is a non-branch instruction or an untaken branch or a taken branch whose target address is the next address.

2. **Intra-cache-line non-sequential flow**
   In this case, *i* is a taken branch instruction and *j* is its target, *i* and *j* reside on the same cache-line and *j* is not the next address of *i*.
3. **Inter-cache-line sequential flow**
   This case is similar to the first one, the only difference is that *i* and *j* reside on different cache-lines.
4. **Inter-cache-line non-sequential flow**
   This is similar to the second case, but *i* and *j* reside on different cache-lines.

In the first case, it is possible to identify the way number for *j* by memoizing the way number for *i*. Therefore, without performing any tag check, we can find the way for instruction *j* [3, 4, 10]. Since most instructions are sequentially executed, this technique successfully reduces the number of tag and way accesses. This technique, however, is not effective for inter-cache-line sequential-flow. Ma et al. [11] proposed a dynamic way-memoization technique which eliminates the way-search operation even for inter-cache-line sequential-flow. They augmented each cache line with a *sequential link*. The sequential link had a field indicating whether the link was valid, and another field pointing to the way of the cache holding the next instruction. They proposed a similar technique for intra and inter cache-line non-sequential flows using branch links instead of sequential links. The downside of the approach is that it requires two extra bits per instruction (one valid bit and one way bit). Therefore, some extra energy is consumed to read the additional bits. Furthermore, they require a mechanism to invalidate sequential and branch links upon a cache-line replacement. Unlike their approach, ours do not need such a mechanism.

Another approach which can handle *non-sequential flow* is based on Branch Target Buffer (BTB) [12]. Inoue et al. extended BTB and used it to reduce the number of tag checks for *non-sequential flow*. Their approach, however, cannot handle the *inter-cache-line sequential flow*. Our approach can handle both *inter-cache-line sequential and non-sequential flow*.

For data caches, Su and Despain proposed in-cache two-level hierarchies in which a single line-buffer is accessed before the main cache [13]. The single line-buffer works as a first level cache. This is conceptually the same as a single-entry filter cache [6]. This approach, however, degrades the performance because a line-buffer miss will require additional cycles to access the main cache.

Yang et al. [14] proposed a lightweight set buffer to exploit *set-wise access locality* in data caches (a set includes the lines in different cache-ways corresponding to the same set-index). No additional cycle is required in case of a set buffer miss. This technique, however, cannot exploit inter-cache-line access locality.

Ghose and Kamble [15] proposed a multiple line-buffer technique in which cache lines in the same set are organized in a single Wide Line Buffer (WLB) and multiple such lines are kept. Each WLB entry has data, tag, and index number fields for each cache line. This technique improves line buffer hit rate, but there is an energy overhead associated with the method due to the power wasted to access the WLB on a WLB miss and the power consumed for accessing set-indices of the WLB for every cache access. Since their technique keeps values of cache lines, its area overhead is very large. Unlike their approach, the overhead of our approach is small (around 3%).

Witchel et al. [16] presented a *direct addressing* scheme which allows software to access cache data directly without tag checks. The idea is to memoize the location of a cache line so that hardware can eliminate tag checks when software access the line again. They employ several *direct address registers (DARs)* which are used by software to memoize the location of cache lines. The main shortcoming of the scheme is the necessity of using special load and store instructions and compiler support for them.

Ashok et al. [17] presented a *Cool-Mem* scheme which keeps several number of recently used addresses in the "Hotline Registers" and skip tag lookups and redundant way accesses when there is a hit in the "Hotline Registers". However, it requires the existence of a TLB access stage between the address generation stage and the cache access stage. Otherwise, the technique requires an extra cycle for the Tag-Cache lookup. Our technique does not suffer from this limitation because the table (MAB) lookup is done in parallel with the address calculation.

## 3 Our Methods

Since the memory address generation unit is on the critical path in many processors, accessing the address generation unit and the MAB sequentially in the same pipeline stage increases the cycle time. To solve this problem, instead of addresses we keep tag and set-index values in the MAB (see Figure 1 and 2) and access them in parallel with the address generation unit. The technique is based on the observation that the target address is the sum of a base address and a displacement which is typically small [18, 19]. To the best of our knowledge, our method is the first one which exploits small displacements in the context of way memoization for data caches. Additionally, for the first time we use the fact that most branch offsets are small to reduce the power consumption of instruction caches.

### 3.1 Way-memoization for D-caches

The base address and the displacement for load and store operations usually take a small number of distinct values [18, 19]. Therefore, we can improve the hit rate of the MAB by keeping only a small number of most recently used tags. Assume the bit width of tag memory, the number of sets in the cache, and the size of cache lines are 18, 512, and 32 bytes, respectively. The width of the set-index and offset fields will be 9 and 5 bits, respectively. Since most (according to our experiments, more than 99% of) displacement values are less than $2^{14}$, we can easily calculate tag values without address generation. This can be done by checking the upper 18 bits of the base address, the sign-extension of the displacement, and the carry bit of a 14-bit adder which adds the low 14 bits of the base address and the displacement. Therefore, the delay of the added circuit is the sum of the delay of the 14-bit adder and the delay of accessing the set-index table. Our experiment shows this delay is smaller than the delay of the 32-bit adder used to calculate the address. Therefore, our technique does not have any delay penalty. Note that if the displacement value is more than or equal to $2^{14}$ or less than $-2^{14}$, there will be a MAB miss, but the chance of this happening is less than 1%. The details of the MAB architecture and synthesis results will be presented in Section 3.3 and Section 4, respectively.

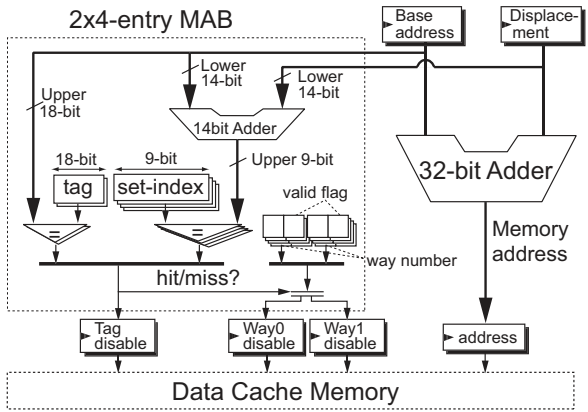

Figure 1: Way memoization for D-cache

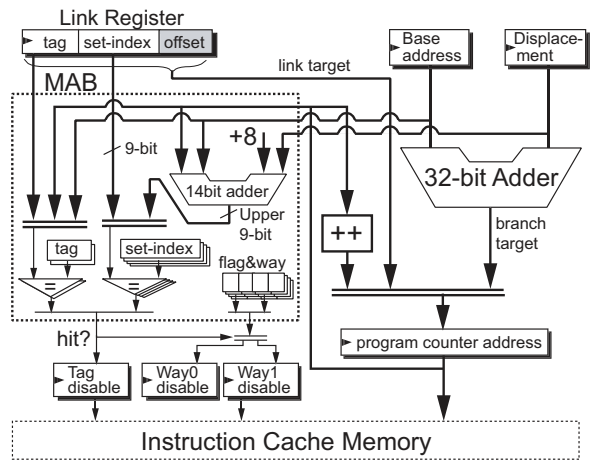

Figure 2: Way memoization for I-cache

### 3.2 Way-memoization for I-caches

To eliminate redundant tag and way accesses for *inter-cache-line flows* (see Subsection 2), we can use a MAB. Unlike the MAB used for D-cache, the inputs of the MAB used for I-cache can be one of the following three types: 1) an address stored in a link register, 2) a base address (i.e., the current program counter address) and a displacement value (i.e., a branch offset), and 3) the current program counter address and its stride. In the case of *inter-cache-line sequential flow*, the current program counter address and the stride of the program counter are chosen as inputs of the MAB. The stride is treated as the displacement value. If the current operation is a *"branch (or jump) to the link target"*, the address in the link register is selected as the input of the MAB as shown in Figure 2. Otherwise, the base address and the displacement are used as done for the data cache.

### 3.3 The MAB architecture

The MAB has two types of entries: 1) tag and cflag (20 bits), 2) set-index (9 bits). The 2-bit cflag is used to store the carry bit of the 14-bit adder and the sign of the displacement value. If the number of entries for tags is $n_1$ and the number of entries for set-indices is $n_2$, the MAB can store the information about $n_1 \times n_2$ addresses. For example, a $2 \times 8$-entry MAB can store information about 16 addresses. For each address, there is a flag indicating whether the information is valid. The flag corresponding to the tag entry $i$ and set-index entry $j$ is denoted by vflag[$i$][$j$]. The MAB entries are updated using Least Recently Used (LRU) policy [20].

Consider an address corresponding to a tag value $x$ and a set-index value $y$. Depending on whether there is a hit or miss for $x$ and $y$, there are four different possibilities,

1. There are hits for both $x$ and $y$. In this case the address corresponding to $(x, y)$ is in the table. Assuming $i$ and $j$ denote the entry numbers for $x$ and $y$, respectively, vflag[$i$][$j$] is set to 1.

2. There is a miss for $x$ and a hit for $y$. If $j$ denotes the entry number for $y$ and $x$ replaces entry $i$ in the MAB, vflag[$i$][$j$] has to be set to 1, while other vflag[$i$][*] are set to 0.

3. There is a hit for $x$ and a miss for $y$. Assuming $i$ denotes the entry number of $x$ and $y$ replaces entry $j$ in the MAB, vflag[$i$][$j$] is set to 1, while other vflag[*][$j$] are set to 0.

4. Finally, there are misses for both $x$ and $y$. If $x$ and $y$ replace entry $i$ and entry $j$ in the MAB, vflag[$i$][$j$] will be set to 1 and other vflag[$i$][*] and vflag[*][$j$] will be set to 0.

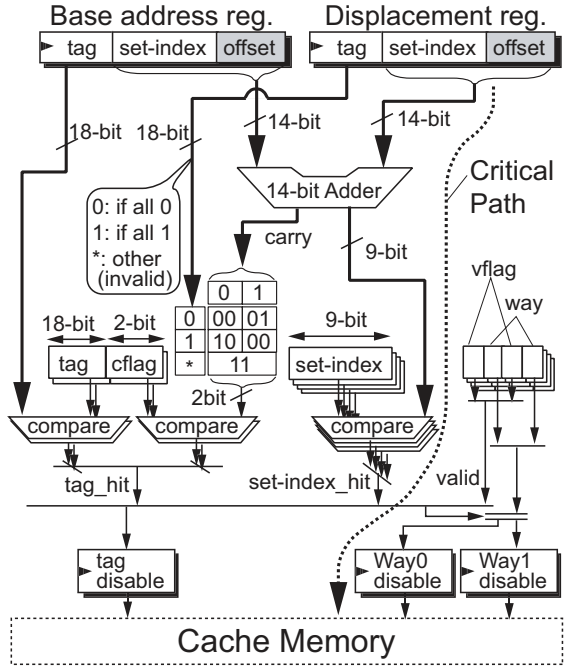

Figure 3: Detailed structure of MAB

To keep the MAB consistent with the cache, if not all upper 18 bits of the displacement are zero and not all of them are one, vflags corresponding to the entry $LRU$ (i.e., vflag[$LRU$][*]) are set to 0. As long as the number of tag entries in the MAB is smaller than the number of cache-ways, this guarantees the consistency between the MAB and the cache. In other words, if a tag and set-index pair residing in the MAB is valid, data corresponding to them will always reside in the cache. Figure 3 shows the details of the MAB. The critical path delay is the sum of the delay of the 14-bit adder and the delay of the 9-bit comparator which is smaller than the clock period of our target processor.

## 4 Experimental results

We applied our technique to Fujitsu VLIW processor (FR-V) [21] designed in a $0.13\mu m$ CMOS process technology with a 1.3V supply voltage and the clock speed of 360MHz. The processor employs two 32kB 2-way set associative caches for instruction and data. The number of sets and cache line size for both caches are 512 and 32 bytes, respectively.

Table 1: Area Overhead ($mm^2$)

|  |  | # entries for set-indices | | | |
|---|---|---|---|---|---|
|  |  | 4 | 8 | 16 | 32 |
| # entries for set-indices | 1 | 0.016 | 0.027 | 0.065 | 0.307 |
|  | 2 | 0.019 | 0.033 | 0.085 | 0.311 |

Table 2: The delay of the added circuit (ns)

|  |  | # entries for set-indices | | | |
|---|---|---|---|---|---|
|  |  | 4 | 8 | 16 | 32 |
| # entries for set-indices | 1 | 1.00 | 1.00 | 1.08 | 1.14 |
|  | 2 | 1.02 | 1.02 | 1.08 | 1.16 |

To evaluate the power, area, and delay overhead of our method, we implemented the MAB circuits in Verilog and synthesized them using SYNOSPSYS Design-Compiler. Table 1 shows the area overhead for different number of entries of the MAB. The tag and the set-index are (18+2)-bit and 9-bit, respectively. Based on our experiments with different benchmark programs, a MAB with two entries for tag and 8 entries for set-index is optimal from the power consumption viewpoint for all application programs we studied. The area overhead of this configuration when used for a D-cache is around 3%. For I-cache depending on the application program, one of $2 \times 16$-entry or $2 \times 32$-entry configurations is optimal; since the former has a smaller area overhead than the latter (7.5% compared to 27.5%), we used the $2 \times 16$-entry configuration for our processor.

The critical path delay of the MAB is the sum of the delay of the 14-bit adder and the delay of the 9-bit set-index comparator as shown in Figure 3. Table 2 shows the delay for different configurations. Since the maximum clock frequency of our target processor is 400MHz [21], CPU cycle time of the processor is 2,500ps. Therefore, the delay of the MAB is much smaller than the CPU cycle time. Since the MAB is accessed in parallel with the 32-bit adder for address generation, our approach does not increase the CPU cycle time. Table 3 shows the power consumption of different MAB configurations. We used SYNOPSYS NanoSim for power estimation. Since we used clock gating in our circuits, the power consumptions were very small when the circuits were not used. We used NanoSim and Softune Ver.6 [22] (the instruction-set simulator of FR-V processor) to estimate the power consumption of caches after modification. The power consumption results include the leakage power. We used seven benchmark programs, DCT, FFT, whetstone, dhrystone, compress, jpeg encoder, and mpeg2 encoder.

Table 3: Power consumption (mW)

|  |  |  | # entries for set-indices | | | |
|---|---|---|---|---|---|---|
|  |  |  | 4 | 8 | 16 | 32 |
| # of entries for set-indices | 1 | active | 1.95 | 2.37 | 3.39 | 6.25 |
|  |  | sleep | 0.24 | 0.40 | 0.76 | 1.37 |
|  | 2 | active | 2.34 | 3.07 | 4.56 | 7.93 |
|  |  | sleep | 0.40 | 0.68 | 1.28 | 2.26 |

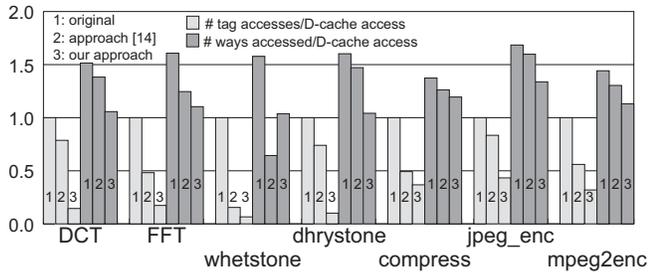

Figure 4: Tag and way accesses for D-cache

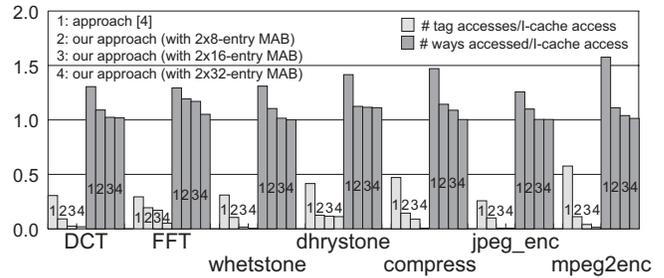

Figure 6: Tag and way accesses for I-cache

Figure 4 shows the average number of tag and way accesses per D-cache access. The processor uses a write-back buffer which makes it possible to access only a single way for store instructions. As a result the number of ways accessed per D-cache access is less than 2 in all cases. Since in our approach at least one way is accessed per cache access, the number of ways accessed per D-cache access is more than 1. On the other hand, the number of tag accesses is reduced by 90% compared to the original cache architecture. Figure 5 shows the power consumption of the D-cache calculated using the following formula,

$$P_{Dcache} = E_{way} \times N_{way} + E_{tag} \times N_{tag} + P_{MAB} \quad (1)$$

where $E_{way}$, $N_{way}$, $E_{tag}$, $N_{tag}$, and $P_{MAB}$ are the energy consumption per cache-way access, the number of ways accessed per second, the energy consumption per tag memory access, the number of tags accessed per second, and the power consumption of the MAB, respectively. $E_{way}$ and $E_{tag}$ were estimated using SPICE. $N_{way}$ and $N_{tag}$ were measured using an instruction-set simulator [22]. $P_{MAB}$ was the power consumption of a $2 \times 8$-entry MAB from Table 3. $N_{way}$ is equal to $N_{load} + N_{store}$, where $N_{load}$ and $N_{store}$ denote the number of ways accessed per second for load and store operations, respectively. Our approach for D-cache does not change $N_{store}$, but it reduces $N_{load}$ and $N_{tag}$. The results in Figure 5 show that our approach reduces the power consumption in D-cache by 35% on an average.

Figure 6 shows the average number of tags and ways accessed per I-cache access. In the case of *intra-cache-line sequential flow*, no tag access is required [3, 4, 10], as the current address is guaranteed to be found in the same cache line as the previous address. The left-most bar for each benchmark program represents the result when this optimization is performed. This optimization reduces the number of tag accesses by 60% on an average. Our approach with a $2 \times 16$-entry MAB reduces the average number of tag accesses to 80% of the approach [4].

Figure 7 shows the power consumption results for I-cache. Our approach with a $2 \times 16$-entry MAB can reduce the power consumption by 25% on an average. Finally, Figure 8 shows the total power consumption of I-cache and D-cache. We used a $2 \times 16$-entry MAB and a $2 \times 8$-entry MAB for I-cache and D-cache, respectively. The total power consumption was reduced on an average by 30%. The maximum saving was 40% achieved for the *mpeg2enc* program.

## 5 Conclusion

We proposed a technique to eliminate redundant cache-tag and cache-way accesses to reduce power consumption. We applied the proposed technique to FR-V processor and estimated its power saving using NanoSim and ISS (Softune Ver.6) [22]. Our experiments for 32kB 2-way set asso-ciative caches show the power consumption can be reduced by up to 40% (30% on an average) without any performance penalty. We are currently extending our approach by combining it with the line buffer technique to achieve more saving.

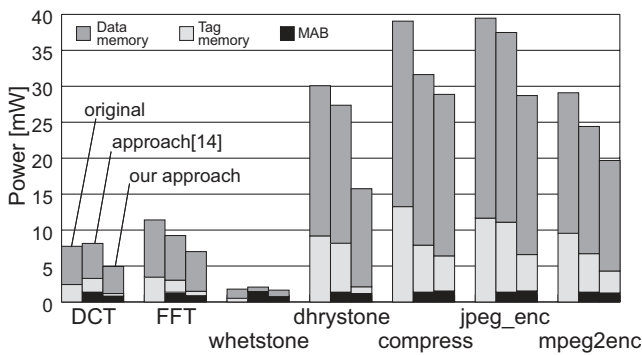

Figure 5: Power consumption results for D-cache

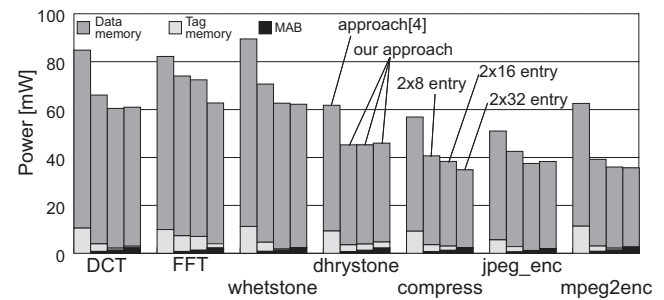

Figure 7: Power consumption results for I-cache

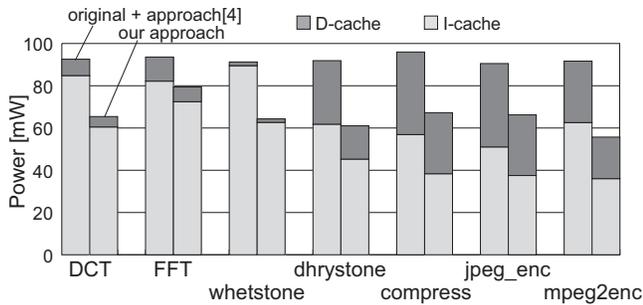

Figure 8: Total power consumption

## Acknowledgments

We appreciate Takashi Shikata, Naoshi Higaki, Takao Sukemura and their group at Fujitsu Ltd., Yutaka Tamiya, Koichiro Takayama, and Kaoru Kawamura at Fujitsu Labs. Ltd., and Tom Sidle at Fujitsu Labs. of America, Inc. for assisting us in performing this research.